\documentclass[twocolumn,prb,aps,epsf,showpacs]{revtex4-1}

\usepackage{graphicx}
\usepackage{dcolumn}
\usepackage{bm}

\begin{document}

\title{Resistivity saturation in a weakly interacting 2D Fermi liquid at intermediate temperatures}

\author{Xiaoqing Zhou$^{1}$, B. Schmidt$^{1}$, L.W. Engel$^{2}$, G. Gervais$^{1}$, L.N. Pfeiffer$^{3}$ and K.W. West$^{3}$, S. Das Sarma$^{4}$,}

\affiliation{$^{1}$ Department of Physics, McGill University,
Montreal, H3A 2T8, CANADA}

\affiliation{$^{2}$ National High Magnetic Field Laboratory, Tallahassee, FL 32310, USA}

\affiliation{$^{3}$ Department of Electrical Engineering, Princeton University, Princeton NJ 08544 USA}

\affiliation{$^{4}$ Condensed Matter Theory Center, Department of
Physics, University of Maryland, College Park, MD 20742 USA}

\date{\today }

\begin{abstract}
We report a highly unusual temperature dependence in the magnetoresistance of a weakly interacting high mobility 2D electron gas (2DEG) under a parallel magnetic field and in the current configuration $\textbf{\emph{I}} \perp\textbf{\emph{B}}$. While the linear temperature dependence below 10~K and the exponential temperature dependence above 40~K agree with existing theory of electron-phonon scattering, a field induced resistivity saturation behaviour characterized by an almost complete suppression of the temperature dependence is observed from approximately 20 to 40~K, which is in sharp contrast to the phenomenology observed in the configuration $\textbf{\emph{I}} \parallel \textbf{\emph{B}}$.  Possible origins of this intriguing intermediate temperature phenomenon are discussed. 
\end{abstract}

 \maketitle

The dependence of resistance on temperature is a key experimentally observable feature of electronic systems, and provides an important testing ground for theoretical models of the microscopic scattering mechanisms at play. For the well understood Fermi liquid theory based systems, scattering is usually governed by the low energy excitations near the Fermi surface, and resistance typically decreases as the temperature is lowered since there are fewer states to scatter from and into. Nevertheless, lowering the thermal energy can often allow other effects to become more prominent, some of which can give rise to a negative or more generally non-monotonic temperature dependence. These include disorder-related effects such as Anderson localization\cite{Anderson58PR, Abrahams79PRL}, Coulomb interaction related effects such as Mott physics\cite{Mott}, various spin-related phenomena and other interaction-driven quantum phase transitions, all of which require working at the lowest possible temperatures. In contrast, the intermediate temperature regime from $\sim$4~K to $\sim$77~K has been relatively neglected. This is perhaps not surprising, since at a sufficiently high temperature the dominating electron-phonon scattering typically results in a strong positive temperature dependence, which has been well understood in conventional metals over a century ago, and in 2DEG over 20 years ago\cite{Sankar90PRB, Sankar92PRB}. Therefore, it is highly unusual and counterintuitive to observe a nearly flat temperature dependence in the intermediate temperature regime in a Fermi liquid system, which is what we report in this work.

The application of a magnetic field brings another dimension of possibility to the electrical transport properties. While applying a magnetic field perpendicular to a 2D system leads to the extensively studied quantum Hall physics\cite{Klitzing, StormerTsui}, here we focus on the relatively unexplored case when the field is applied parallel to the 2D conducting plane. In a typical quantum well 2DEG sample, the parallel field induced quasiparticle cyclotron motion is largely forbidden by the confinement potential, and it is often assumed that the most important effect of the parallel field is spin polarization\cite{Pudalov97JETP, Simonian97PRL, Yoon00PRL}. However, for a 2DEG with a large quantum well width, the coupling between the field and the $\hat{z}$ degree of freedom of the quasiparticle motion can be highly nontrivial when the magnetic length $l_{{\parallel}}\equiv \sqrt{\frac{\hbar c}{eB_{\parallel}} }$ becomes comparable to the quasiparticle confinement width $d_{z}\equiv \sqrt{<z^{2}>}$, giving rise to mixing of 2D subbands. In a previous study\cite{Zhou10PRL}, in the low temperature limit ($<1~$K) and in both the current configurations $\textbf{\emph{I}} \parallel\textbf{\emph{B}}$ and $\textbf{\emph{I}} \perp\textbf{\emph{B}}$, the impact of the magneto-orbital coupling has been observed as a colossal magnetoresistance (CMR) with two changes in slope in the vicinity of 9~T, where the condition $l_{\parallel}\approx d_z$ is met. This is in good agreement with the magneto-orbital coupling theoretical arguments, suggesting that the 2D system gradually evolves into a novel quasi-3D phase in this field range, as different 2D subbands become strongly coupled.


The magneto-orbital coupling has a profound and unexpected impact on the temperature dependence of the 2DEG in the electron-phonon scattering regime. Naively, the $\textbf{\emph{I}} \parallel\textbf{\emph{B}}$ case might be expected to be relatively simple, since the Hamiltonian $\frac{1}{2m^*}(\hat{p}-eA)^2$ has no field term along the current direction; we nevertheless find an apparent field-induced metal-insulator transition phenomenology characterized by a sign change in the temperature slope $\frac{\partial \rho_{xx}}{\partial T}$ below $\sim$10~K\cite{Zhou11MIT}.  The ``transition" occurs within a narrow field window near $\sim 10.5~$T instead of  a single critical field, and coincides with the 2D to quasi-3D crossover transition. This has been interpreted as a quantum-classical crossover effect\cite{Sankar99PRL, Sankar00PRL}. The negative temperature slope is not related to a true ``insulator" phase (in fact, the metallicity parameter $k_F \lambda$ remains large at high fields, $\lambda$ being the mean free path), but an indication of a field-suppressed Fermi temperature $T_F(B) \ll T$ in the quasi-3D phase. We note that this ``classical" regime is accessible in conventional metals ($T_F \sim 10^5$~K) only in the plasma phase. By contrast, in our case the electron-phonon scattering is still dominant at an intermediate temperature range ($>10~$K), signalling a novel regime for theoretical understanding and experimental exploration. 


In this work, we explore the temperature dependence in the $\textbf{\emph{I}} \perp\textbf{\emph{B}}$ case on the same 2DEG sample. Surprisingly, the phenomenology here is distinctively different from that in the $\textbf{\emph{I}} \parallel\textbf{\emph{B}}$ case. It does not show the negative temperature slope below $\sim$10~K at all. Instead, the system remains largely ``metallic" ($\frac{\partial \rho_{xx}}{\partial T}>0$) below $\sim$20~K and above $\sim$40~K, while in the intermediate range the temperature dependence of resistance is largely suppressed by a parallel magnetic field as low as 5 T. The distinct phenomenologies, occurring at field and temperature ranges that differ by about a factor of two, suggest that the contrasting behaviours may result from entirely different mechanisms, although both could ultimately be driven by the same magneto-orbital coupling. The occurrence of these contrasting and distinct phenomenologies in a nearly ideal Fermi gas system poses an intriguing theoretical challenge. The flat temperature dependent resistivity observed in the intermediate ($\sim$20 - 40~K) temperature range is somewhat reminiscent of the resistivity saturation in metals for $T >600~$K typically\cite{Han03RMP}. 


The sample used in this study is a typical AlGaAs/GaAs/AlGaAs 2D system with an ultra-high mobility of $\mu\simeq 10^{7}$~$cm^{2}/V\cdot s$. It is chosen with a charge carrier density of  $n\simeq 10^{11}$ $cm^{-2}$  so that the spin polarization effect is very weak ($E_z/E_{F}$ less than $10~\%$ at 10~T). The electron-electron interaction energy $E_{ee}$ is also comparable to the Fermi energy $E_F$ ($r_s = \frac{E_{ee}}{E_F}$ is about 1.8 in 2D and 1 in quasi-3D), allowing the system to be described as a weakly interacting Fermi liquid. Given the very high mobility, disorder and localization effects should not play a significant role. In such a high mobility sample, we expect the resistivity behaviour in a magnetic field to be dominated by the modification of the fermi surface of the 2DEG, and its interaction with the phonon bath. The well width is chosen to be 40~nm so that the magneto-orbital coupling effect is strong\cite{Zhou10PRL}, and only a single subband is occupied at zero field.  The sample has a rectangular shape with a long-to-short axis ratio $\sim$ 3:1. Except where noted, both the longitudinal resistance $R_{xx}$ and the Hall resistance $R_{xy}$ were measured in the rectangular Van der Pauw geometry using a standard low frequency (13.5~Hz) lock-in technique, with 100~nA current being applied along the long axis. Using an {\it in situ} rotation stage, the 2D electron plane of the sample was carefully aligned to be parallel to the applied magnetic field by nulling the Hall resistance. The qualitative behavior of $R_{xx}(B, T)$ was confirmed to be insensitive to possible angular systematic errors. Its robustness against possible field and temperature errors was also verified by reproducing the data in three facilities with distinctively different field, temperature control and thermometry settings. Unless otherwise stated, the data we present here were taken in the NHMFL hybrid magnet facility, which supplied magnetic fields from 0 to 45~T. 

\begin{figure}[hbt]
\includegraphics[width=0.8\linewidth,angle=0,clip]{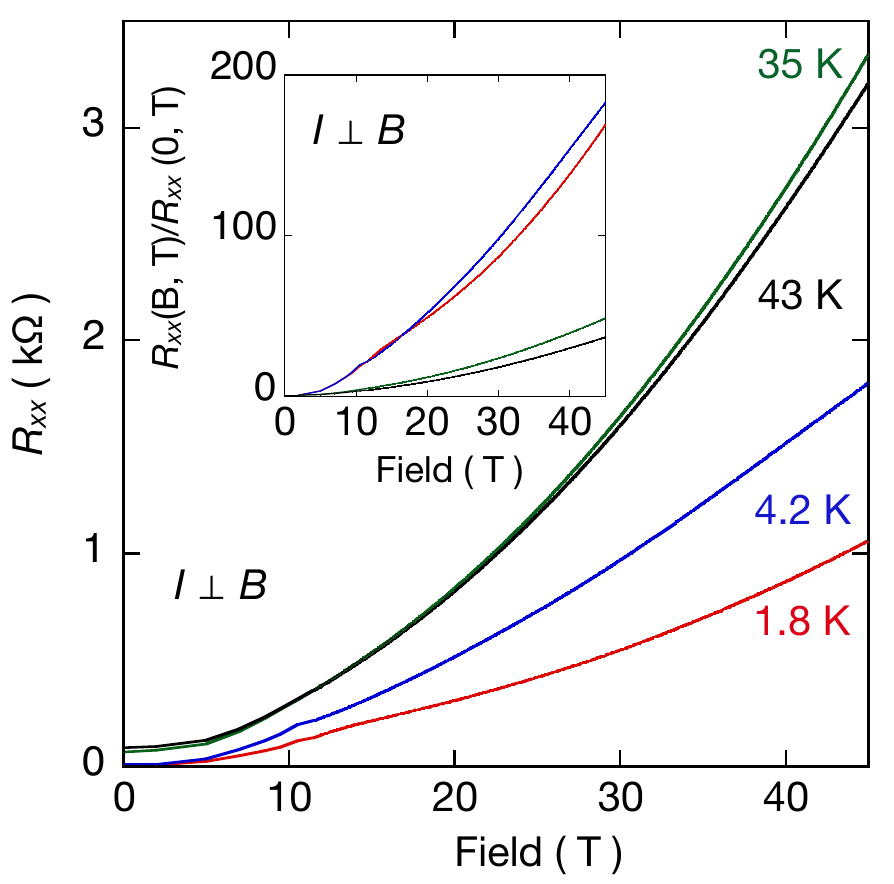}
\caption{ (Color online) Examples of longitudinal resistance $R_{xx}$ as a function of field at several temperatures in the configuration $\textbf{\emph{I}} \perp \textbf{\emph{B}}$. Inset: the normalized resistance as a function of field at several temperatures (matching those in the main panel)  in the configuration $\textbf{\emph{I}} \perp \textbf{\emph{B}}$.} 
\label{fig2}
\end{figure}

The field dependence of $R_{xx}$ is shown in Fig.~\ref{fig2}. At all temperatures from 1.8~K to 43~K, the resistance increases with an increasing field. The low temperature ($<4.2$~K) data are consistent with that of a previous low temperature ($<1$~K) study\cite{Zhou10PRL}. With an increasing temperature, the qualitative shape of the magnetoresistance remains largely unchanged. 

\begin{figure}[h]
\includegraphics[width=1.0\linewidth,angle=0,clip]{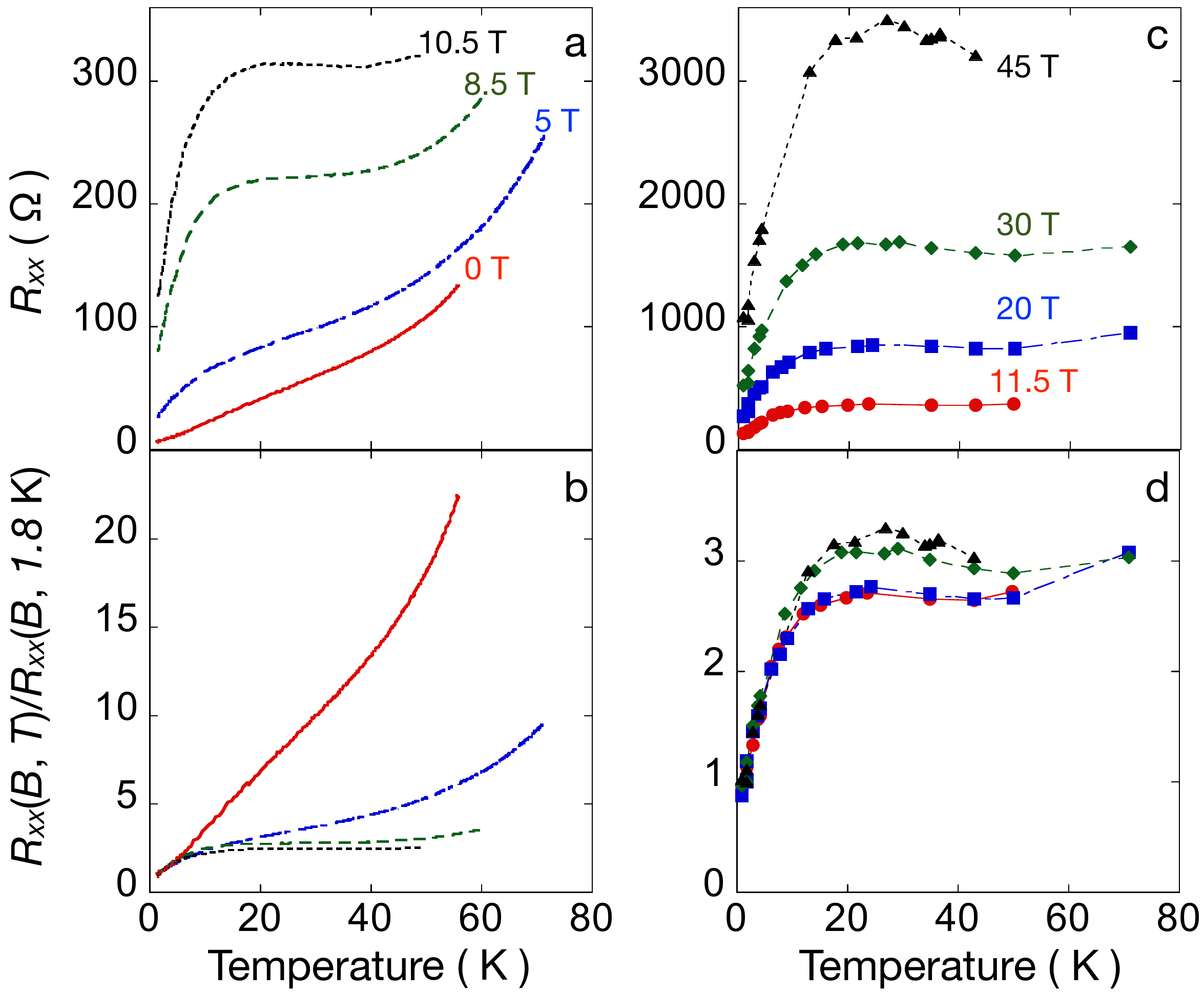}
\caption{(Color online) a) Longitudinal resistance $R_{xx} (B, T)$ and b) normalized resistance from 0~T to 10.5~T in the configuration $\textbf{\emph{I}} \perp \textbf{\emph{B}}$.  c)  Longitudinal resistance $R_{xx} (B, T)$ and d) normalized resistance from 11.5~T to 45~T in the configuration $\textbf{\emph{I}} \perp \textbf{\emph{B}}$. In both the low temperature and high temperature limits the qualitative temperature dependence remains qualitatively unsuppressed by the applied magnetic field, but in the intermediate region a resistivity saturation is observed.} 
\label{fig1}
\end{figure}

Our main finding, a resistivity saturation behaviour in the intermediate ($\sim$20 - 40~K) temperature regime before optical phonon scattering becomes important, is presented in Fig.~\ref{fig1}. Fig.~\ref{fig1}a shows examples of the low field temperature sweep data taken at 0~T, 5~T, 8.5~T and 10.5~T. At zero field, the resistance increases linearly with temperature, while above $\sim$40~K an exponential increase becomes apparent. These temperature dependences are well described by an established theory\cite{Sankar90PRB, Sankar92PRB} as the signatures of the electron-acoustic-phonon scattering and electron-optical-phonon scattering respectively. However, starting from 5~T, we observe the development of an apparent suppression of the temperature dependence in an intermediate temperature range from approximately 10 to 40~K. This suppression becomes stronger as the magnetic field increases, and at $\sim$8.5~T develops into an astonishing $\frac{\partial \rho_{xx}}{\partial T}\sim 0$ plateau that covers a large intermediate temperature range from about 20 to 40~K. Above $\sim$9~T, $\frac{\partial \rho_{xx}}{\partial T}$ gradually becomes weakly negative with an increasing field. Above 11.5~T, $R_{xx}(T)$ at different fields were extracted from the field sweep data taken at several fixed temperatures, but similar resistance saturation behaviours were still observed, as shown in Fig.~\ref{fig1}c. In fact, the weakly negative slope of $R_{xx}(T)$ from 20~K to 40~K persists to the highest field of 45~T, with both its upper and lower boundaries showing only a weak field dependence. On the other hand, the linear temperature dependence at low temperature ($<10~$K) and the exponential temperature dependence at high temperature ($>40~$K) persist from zero field to high magnetic fields, as suggested by the normalized resistances shown in Fig.~\ref{fig1}b and Fig.~\ref{fig1}d. In both the low field ($<11$~T) and high field ($>11$~T) data, below $\sim$10~K the normalized resistivity curves all join a single trend, implying that the acoustic phonon scattering remains robust against a strong magnetic field, and that the resistance is simply scaled by a field-dependent factor. A similar argument can also be applied to the optical phonon scattering, as the exponential temperature dependence beginning at $>40~$K remains apparent at all magnetic fields. If this exponentially increasing contribution of optical phonon scattering is subtracted out from our measured resistivity, then the flat region is extended much further into the higher temperature regime. Therefore, the resistance saturation in the intermediate temperature within the phonon scattering regime is truly puzzling.


\begin{figure}[hbt]
\includegraphics[width=1.0\linewidth,angle=0,clip]{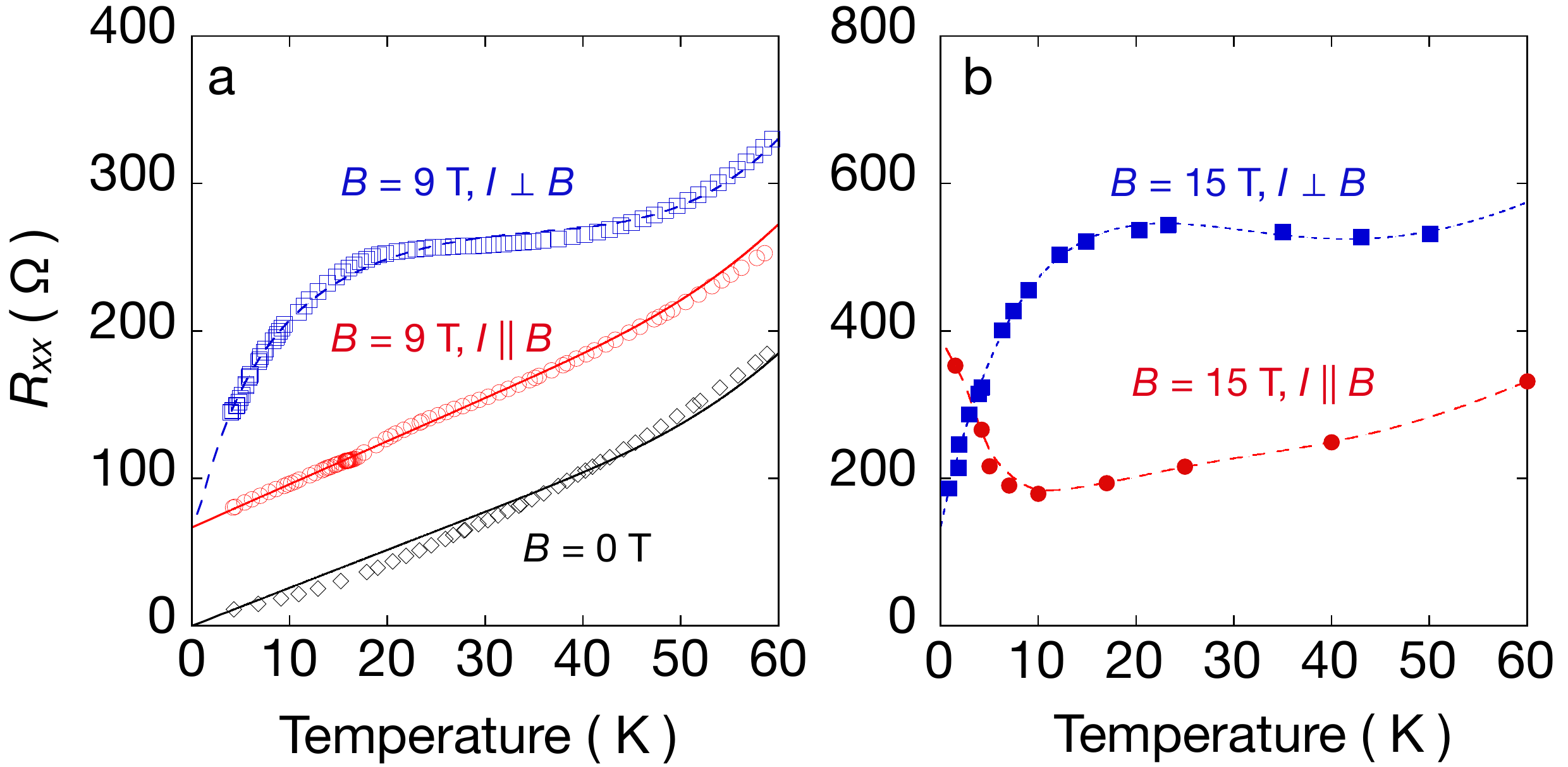}
\caption{(Color online) a) The resistance anisotropy data at 0~T (black square); 9~T with  $\textbf{\emph{I}} \parallel \textbf{\emph{B}}$ (red circle); 9~T with  $\textbf{\emph{I}} \perp \textbf{\emph{B}}$ (blue square), extracted from the temperature sweep data at fixed fields. The resistivity saturation effect is only apparent when  $\textbf{\emph{I}} \perp \textbf{\emph{B}}$. b) The resistance anisotropy data at 15~T with  $\textbf{\emph{I}} \parallel \textbf{\emph{B}}$ (red circles); 15~T with  $\textbf{\emph{I}} \perp \textbf{\emph{B}}$ (blue square), extracted from the field sweep data at fixed temperatures. The 0~T and the 9~T  $\textbf{\emph{I}} \parallel \textbf{\emph{B}}$ data can be fit to Eq.1 with $a_1=0~\Omega$, $a_2=2.57~ \Omega$K, $b_1=3.45\times10^6~ \Omega$K and $b_2=453$K (black solid line) and $a_1= 66.82~\Omega$, $a_2=2.92~\Omega$K, $b_1=3.45\times10^6~\Omega$K and $b_2=453$K (red solid line). The others (9~T $\textbf{\emph{I}} \perp \textbf{\emph{B}}$ data, 15~T $\textbf{\emph{I}} \perp \textbf{\emph{B}}$ data and 15~T $\textbf{\emph{I}} \parallel \textbf{\emph{B}}$ data) cannot be fit to this functional form, and the dashed lines are guides-to-the-eye.} 
\label{fig3}
\end{figure}

To shine more light into this puzzling phenomenology, we compare it with that in the better understood $\textbf{\emph{I}} \parallel \textbf{\emph{B}}$ case, as shown by the examples in Fig.~\ref{fig3}. When $T \rightarrow~0$, the qualitative field dependence of the magnetoresistance was found to be relatively independent of the current configuration\cite{Zhou10PRL}. However, the magnetoresistance at higher temperatures was found to be increasingly anisotropic. In the $\textbf{\emph{I}} \parallel \textbf{\emph{B}}$ case, the resistivity saturation effect is absent, and below the 2D to quasi-3D transition($\sim$ 10.5~T) the temperature dependence is well described by the electron-phonon scattering\cite{Sankar90PRB, Sankar92PRB}:
\begin{equation}
R_{xx} = a_1+a_2 T+b_1 \frac{e^{-b_2/T}}{T},
\end{equation}
\noindent where $a_1+a_2T$ is the acoustic phonon term, and $b_2=\hbar \omega/ k_B$ is the effective optical phonon energy. As shown in Fig.~\ref{fig3}a (with data taken in the McGill facility), fitting parameter $a_1(B)$ seems to be simply the colossal magnetoresistance $R_{xx}(B, 0)$ observed in the low temperature limits. The fact that $a_2$ and $b_1$ do not show a strong field dependence suggests that the phonon scattering is largely unaffected by the applied field. This is perhaps not surprising, as the quasiparticles moving along the field direction should not ``see" the magnetic field, but only the field-deformed Fermi surface. On the contrary, in the $\textbf{\emph{I}} \perp \textbf{\emph{B}}$ case, below $\sim$10~K the linear (acoustic phonon scattering) temperature slope seems to be enhanced by a field-dependent factor. One simple possibility is that the effective mass $m^*$ increases with an increasing field, resulting in an enhanced Drude resistivity $\rho \sim m^*$.\cite{Smrchka95JPC} However, a single field-dependent factor cannot explain the weakly temperature dependent ``plateau" above $\sim$20~K, which persists over a large field range, and the lower bound of which ($< 5~$T) is well below the 2D to quasi-3D crossover transition $\sim 10.5~$T. In fact, considering the strong field dependence of the linear temperature slope below 10~K, it is surprising that the temperature dependence in the resistivity saturation plateau is only very weakly field dependent. This might imply that there is another effect that almost exactly cancels the enhanced acoustic phonon scattering in a large temperature window.  


The anisotropy persists to very strong fields above the the 2D to quasi-3D transition($\sim$ 10.5~T), albeit in a different way. As shown in Fig.~\ref{fig3}b, in the $\textbf{\emph{I}} \parallel \textbf{\emph{B}}$ case (with data taken in the Toulouse pulse field facility), there is an emerging negative temperature slope below 10~K, reminiscent of a metal-insulator transition\cite{Zhou11MIT}. This behaviour has been interpreted as the result of a field suppressed Fermi energy $T_F(B)$, as in the classical $T\gg T_F$ regime an upturn $R_{xx} \propto T_F/T$\cite{DasSarma2005} would overcome the positive temperature slope induced by electron-phonon scattering. However, this upturn is clearly absent in the $\textbf{\emph{I}} \perp \textbf{\emph{B}}$ case (with data taken in the NHFML hybrid magnet facility). In fact, the $R_{xx}(B, T)$ curves in these two configurations show contrasting temperature dependence and cross each other below $\sim$ 10~K. 



To summarize, the details of the phenomenology, and in particular the relevant energy scales, are markedly different in these two current configurations. The only resemblance between these two cases is that there is still an evolution from positive to negative temperature slope in the resistivity saturation regime. One possible interpretation could still be a cancellation between the quantum-classical crossover induced resistivity going as $\rho\sim 1/T$\cite{Sankar00PRB} and the acoustic phonon induced resistivity going as $\rho\sim T$. But, given that these two effects have completely independent physical origins, it is difficult to see such a perfect cancellation over a large temperature range. Furthermore, the interpretation of the anisotropy and the apparent energy scale differences is of great theoretical challenge. We also note that common mechanisms that could lead to a negative temperature slope such as the spin degree-of-freedom or Coulomb interactions, can most likely be ruled out in such a high density and high mobility sample. The cutoff temperature as high as $\sim$20~K also poses a challenge to common interpretations. Even the magneto-orbital coupling effect, which has been successful in the $\textbf{\emph{I}} \parallel \textbf{\emph{B}}$ case, has difficulties explaining this phenomenology. Although in the the quantum Hall physics case (where  $\textbf{\emph{I}} \perp \textbf{\emph{B}}$ is always satisfied) field induced change in the functional power law of the electron-phonon scattering term has been reported before\cite{Stormer90PRB}, to the best of our knowledge, a complete suppression of the overall temperature dependence has never been reported. 

It is worth mentioning that certain aspects of the resistivity saturation phenomenology might have been observed in other systems under very different conditions. In particular, a somewhat similar phenomenology was observed in the work by Gao et al.\cite{Gao02PRL} below 1~K,  with a much narrower temperature range and at a lower field range. Similarly, an evolution of a  non-monotonic temperature dependence with the carrier density was reported below 1~K\cite{Millis99PRL}. There are also a few zero field studies with fixed carrier density that perhaps show certain phenomenological resemblances to the resistivity saturation behaviour at intermediate temperature ranges (Fig.~2 in ref. \cite{Spivak10RMP}). We emphasize, however, that none of these studies reported a flat temperature dependence in resistivity over a large temperature range as we have discovered. Furthermore, our system behaves like a conventional 2D metal below $\sim$5~T, whereas the non-monotonic temperature dependence in these other studies already occurs at zero field. Also, the resemblances in certain aspects of the phenomenologies by no means guarantee a similar origin. Given that there are so many differences between these earlier works and ours (external physical conditions, intrinsic physical parameters, details of phenomenology, etc.), at this stage any possible connection between them remains completely unclear.

In conclusion, we have observed the development of a resistivity saturation effect in a temperature range from $\sim$20 to $\sim$40~K, driven by a parallel magnetic field from as low as 5~T to as high as 45~T. The strong suppression of the temperature dependence of resistivity in a well established metallic system, in a wide and high temperature range by a relatively small field, is unprecedented, unexpected and might not be readily understood within the current theoretical framework. We believe that we have discovered a new transport phenomenon in an intermediate temperature regime. 


This work has been supported by NSERC, CIFAR, FQRNT and Microsoft Station-Q. The work at Princeton was partially funded by the Gordon and Betty Moore Foundation as well as the National Science Foundation MRSEC Program through the Princeton Center for Complex Materials (DMR-0819860). A portion of this work was performed at the National High Magnetic
Field Laboratory, which is supported by NSF Cooperative Agreement
No. DMR-0084173, by the State of Florida, and by the DOE. We also thank T. Murphy, E. Palm, R. Talbot, R. Gagnon and J. Smeros for technical assistance.


\begin{thebibliography}{70}


\bibitem{Anderson58PR} P. W. Anderson, Phys. Rev. {\bf 109}, 1492 (1958).

\bibitem{Abrahams79PRL} E. Abrahams {\it et al.}, Phys. Rev. Lett. {\bf 42}, 673 (1979).

\bibitem{Mott} N. F. Mott, Proc. Phys. Soc.  A{\bf 62}, 416 (1949).

\bibitem{Sankar90PRB} T. Kawamura and S. Das Sarma, Phys. Rev. B{\bf 42}, 3725 (1990).

\bibitem{Sankar92PRB} T. Kawamura and S. Das Sarma, Phys. Rev. B{\bf 45}, 3612 (1992).

\bibitem{Klitzing} K. von Klitzing, G. Dorda, and M. Pepper, Phys. Rev. Lett. {\bf 45}, 494 (1980).
 
\bibitem{StormerTsui} D.~C. Tsui, H.~L. Stormer, and A.~C. Gossard, Phys. Rev. Lett. {\bf 48}, 1559 (1982).

\bibitem{Pudalov97JETP} V. M. Pudalov {\it et al.}, JETP Lett. {\bf 65}, 932 (1997).

\bibitem{Simonian97PRL} D. Simonian {\it et al.}, Phys. Rev. Lett. {\bf 79}, 2304 (1997).

\bibitem{Yoon00PRL} J. Yoon {\it et al.}, Phys. Rev. Lett. {\bf 84}, 4421 (2000).

\bibitem{Zhou10PRL} Xiaoqing Zhou {\it et al.}, Phys. Rev. Lett. {\bf 104}, 216801 (2010). 

\bibitem{Zhou11MIT} Xiaoqing Zhou {\it et al.}, Phys. Rev. Lett. {\bf107}, 086804 (2011).

 
 \bibitem{Sankar99PRL} S. Das Sarma and E. H. Hwang, Phys. Rev. Lett. {\bf 83}, 164 (1999).
 
 \bibitem{Sankar00PRL} E. H. Hwang and S. Das Sarma, Phys. Rev. Lett. {\bf 84}, 5596 (2000).
 
 \bibitem{Han03RMP} O. Gunnarsson, M. Calandra and J. E. Han, Rev. Mod. Phys. {\bf 75}, 1085 (2003)
 
 \bibitem{Smrchka95JPC} L. Smr\v{c}ka and T. Jungwirth, J. Phys. Cond. Matt. {\bf 6}, 55 (1994) 
 
 \bibitem{DasSarma2005} S. Das~Sarma and E.~H. Hwang, Phys. Rev. B {\bf 72}, 205303 (2005); {\bf 72}, 035311 (2005), and references therein.
 
 \bibitem{Sankar00PRB} S. Das Sarma and E. H. Hwang, Phys. Rev. B {\bf 61}, R7838 (2000).
 
 \bibitem{Stormer90PRB} H. L. Stormer {\it et al.},  Phys. Rev. B {\bf 41}, 1278 (1990).
 
 \bibitem{Gao02PRL} Xuan P.A. Gao {\it et al.},  Phys. Rev. Lett. {\bf 88}, 166803 (2002).
 
 \bibitem{Millis99PRL} A. P. Millis {\it et al.},  Phys. Rev. Lett. {\bf 83}, 2805 (1999).
 
 \bibitem{Spivak10RMP} B. Spivak {\it et al.}, Rev. Mod. Phys. {\bf 82}, 1743 (2010). 
 
 \end{thebibliography}
\end{document}